\documentclass[preprintnumbers,11pt]{article}
\usepackage{authblk}
\usepackage{amstext,amsmath,amsfonts,amssymb,multicol}
\usepackage{epsfig}
\usepackage{hyperref}
\hypersetup{colorlinks=true,linkcolor=red,citecolor=blue}
\usepackage{graphics}
\usepackage{cite}                                      

\renewenvironment{thebibliography}[1]{
  \begin{oldthebibliography}{#1}
    \setlength{\itemsep}{0em}
    \setlength{\parskip}{0em}
}
{
  \end{oldthebibliography}
}
\providecommand{\keywords}[1]
{
  \small	
  \textbf{\textit{~~~~~~Keywords}:} #1
}
\begin{document}
\title{Late time cosmological solutions in $f(R,T)$ gravity in a viscous Universe}
\author[1]{Hamid Shabani,\thanks{h.shabani@phys.usb.ac.ir}}
\author[2]{Norman Cruz,\thanks{norman.cruz@usach.cl}}
\author[3]{Amir Hadi Ziaie,\thanks{ah.ziaie@riaam.ac.ir}}

\affil[1]{Physics Department, Faculty of Sciences, University of Sistan and Baluchestan, Zahedan, Iran}
\affil[2]{Departamento de F\'isica, Universidad de Santiago de Chile, Avenida Ecuador 3493, Santiago, Chile}\affil[2]{Center for Interdisciplinary Research in Astrophysics and Space Exploration (CIRAS), Universidad de Santiago de Chile, Avenida Libertador Bernardo O’Higgins 3363, Estaci\'on Central, Chile}
\affil[3]{Research Institute for Astronomy and Astrophysics of  Maragha, University of Maragheh,  P. O. Box 55136-553, Maragheh, Iran}

\maketitle
\begin{abstract}
Considering the condition on conservation of energy momentum tensor (EMT), we study late time cosmological solutions in the context of $f(R,T)=R+\alpha T^{n}$ gravity (where $\alpha$ and $n$ are constants) in a flat FLRW spacetime. The present model discusses the case of a barotropic perfect fluid as the matter content of the Universe, along with the case when dissipative effects are taken into account. Briefly, assuming a single perfect fluid we find that  the mentioned model is not capable of presenting an observationally consistent picture of the late time accelerated expansion of the Universe; nevertheless, the model leads to admissible solutions when the bulk viscosity is included. In this regard, a consistent setting is found considering the Eckart, Truncated and Full Israel-Stewart theories which determine the behavior of bulk viscosity. In the presence of bulk viscosity, the behavior of the deceleration parameter (DP) shows that the underlying model can describe an acceptable evolution even if the isotropic pressure of matter content of the Universe is negligible.
\end{abstract}

\vspace{35pt}\tableofcontents
\section{Introduction}\label{int}
During the past decades many attempts have been performed to justify the presently observed accelerated expansion of the Universe. Recent astronomical data have always been in favor of a mysterious repulsive force, which via exerting a negative pressure, leads to accelerated expansion of the Universe. For example observations of Type Ia supernovae~\cite{riess1998,schmidt1998,garnavich19981,garnavich19982,perlmutter1997,perlmutter1998,perlmutter1999}, CMB~\cite{bernardis2000,spergel2003}, baryon acoustic oscillations (BAO)~\cite{eisenstein2005,percival2007} and gravitational lensing~\cite{suyu2010,schrabback2010} probes have measured the accelerated expansion of the Universe, precisely. This phenomenon is attributed to a mysterious energy component of the Universe dubbed dark energy~(DE), see Ref.~\cite{huterer2018} for a recent review. Since, the theory of General Relativity (GR) cannot explain DE within its original version, one usually has two choices; adding a positive cosmological constant (CC) which is denoted by $\Lambda$ or modifying the field equations of GR, at the classical level. The former is known as the $\Lambda {\rm CDM}$ model which also includes a zero pressure cosmic fluid, the so-called dark matter~(DM). DM is responsible for the ancient process of structure formation which has led to the formation of galaxies and galaxy clusters as well as the flatness of galaxies rotation curve~\cite{martino2020,bertone2018}. Recent estimation of proportions of these two exotic components are about $32\%$ for DM as well as baryonic matter and about $68\%$ for DE out of the total energy budget of the Universe~\cite{aghanim2020}.
\par
Since its advent, the $\Lambda {\rm CDM}$ model has gained remarkable achievements, although it suffers from two famous problems: the fine tuning problem, also called the CC problem, which implies that $\Lambda {\rm CDM}$ cannot present a well matched origin for the cosmological constant through the quantum field theory~\cite{zeldovich1968,weinberg1989} and the coincidence problem which poses this question that why the Universe has accelerated just not long ago or long after the present epoch~\cite{carroll2001}. As mentioned above, modifying GR is a promising alternative way to tackle the issue of present day acceleration of the Universe. Among a sea of ideas for modifying GR with the aim of explaining late time behavior of the Universe, T. Harko and his collaborators brought forward this suggestion that incorporating the contribution due to the matter field within the geometry in a novel way may resolve the DE problem. In this regard, they introduced $f(R,T)$ gravity~\cite{harko2011} which involves an arbitrary function of the Ricci scalar $R$ and the trace of EMT instead of the mere presence of Ricci scalar in the gravitational action. This theory has been widely investigated in different areas up to the present time and some interesting subjects of study are: dynamics of scalar perturbations~\cite{alvarenga2013}, late time solutions via the dynamical system approach for a pressureless perfect fluid~\cite{shabani2013,shabani2014}, thermodynamic point of view~\cite{harko2014}, effects of collisional matter on the evolution of the Universe~\cite{baffou2015}, gravitational waves~\cite{alves2016}, the issue of DM~\cite{zaregonbadi2016}, Einstein static solutions~\cite{shabani20171}, consequences of energy conservation violation~\cite{shabani20172}, late time cosmology of $f(R,T)=g(R)+h(T)$~\cite{shabani20173}, wormhole solutions~\cite{moraes2017,elizalde2018}, the unimodular approach~\cite{rajabi2017}, bouncing cosmology~\cite{singh2018, shabani20182}, metric-affine $f(R,T)$ gravity~\cite{barrientos2018}, constrains from Earth’s studies~\cite{ordines2019}, galaxy rotation curves~\cite{shabani2023}, implementation of the torsion field~\cite{saridakis2020} and derivation of the junction conditions~\cite{rosa2021}. The $f(R,T)$ gravity theories have usually been discussed in the literature, disregarding the conservation of EMT. For instance, in Refs.~\cite{singh2014,moraes2015,yadav2015,moraes2018} the model with $f(R,T)=R+\gamma T$ has been utilized where $\gamma$ is a constant. Models of the form $f(R,T)=f(R)+f(T)$ have been investigated in~\cite{baffou2017} with $f(R)=R-2\Lambda(1-\exp(R/b\Lambda))$ and $f(T)=T^{\beta}$, and in~\cite{moraes2019} with $f(R)=R+\alpha R^{2}$ and $f(T)=2\gamma T$. Also, the model with $f(R,T)=R+\alpha T+\beta T^{2}$ has been investigated in~\cite{moraes2016}. The authors of~\cite{sing2014} have discussed the effect of viscosity in $f(R,T)=R+2\alpha T$ theories. In~\cite{shabani20172,shabani20173} two of the present authors have considered cosmological models within $f(R,T)=R+\beta T^{\alpha}$ and $f(R,T)=R^{\alpha}+ T^{\beta}$ formalisms.

\par
Nevertheless, in this paper we study cosmological implications of $f(R,T)=R+\alpha T^{n}$ theory with $n$ and $\alpha$) being constants, in a flat FLRW spacetime filled by a cosmic fluid with a barotropic equation of state (EoS), $p=w\rho$, when the conservation of EMT does matter. In this sense, we will see in Sec.~\ref{sec2} that in the absence of dissipative effects and as long as the conservation of EMT holds, the Universe experiences a phase transition from an initial state with effective\footnote{In modified theories of gravity an effective EoS parameter is defined via $w_{{\rm eff}}=p_{{\rm eff}}/\rho_{{\rm eff}}=-1-2\dot{H}/3H^{2}$ which is related to the deceleration parameter, while, $w$ denotes the EoS parameter of the cosmic fluid via $w=p/\rho$.} EoS $w_{{\rm eff}}=w$, to a final state which can mimic the $\Lambda{\rm CDM}$ era, i.e., whose effective EoS parameter reaches the value $w_{{\rm eff}}=-1+[(w+1)(3 w+1)]/(3 w^2+w+2)$. For a perfect fluid which nearly mimics a cosmic string, i.e., $w\simeq -1/3$ the effective EoS of the Universe approximates that of DE in $\Lambda{\rm CDM}$ model, i.e., $w_{{\rm eff}}\simeq -1$. However, A pressureless perfect fluid gives $w_{{\rm eff}}=-1/2$ which is far from what the astronomical data give us~\cite{planck2018}. This means that $f(R,T)$ gravity in the presence a pressureless cosmic fluid (which is necessary for the formation of large structures) is not capable of describing the present accelerated expansion of the Universe with an admissible accuracy, at least as long as the conservation of EMT is fulfilled. In this regard, to cure the mentioned issue we try to reexamine the problem of late time behavior of the Universe in the EMT conserved version of $f(R,T)$ gravity considering dissipative effects within the cosmic fluid; these  effects arise due to the presence of bulk viscosity in such fluids. The first attempts toward developing a general theory of dissipation in relativistic imperfect fluids was carried out by Eckart~\cite{Eckart919} and in a somewhat different formulation by Landau and Lifshitz~\cite{LanLif}. These theories were further developed by Israel~\cite{Is1976} and Israel and Stewart~\cite{IsStew}. The possibility of explaining late time accelerated expansion of the Universe, as an influence of the effective negative pressure due to bulk viscosity in the cosmic fluids, was first studied in~\cite{zimdahl2001,Pavon2003} and since then several authors have included bulk viscosity in cosmological models~\cite{ribeiro2006,barrow2009,nucamendi2009,nucamendi2010,zimdahl2010,setarsheikh,setarsheikh1,setarsheikh2,sasidharan2015,mathew2016,norman2017,mohan2017, palma2020,hernandez2021,palma2021}. Motivated by these considerations, we here show that in the framework of $f(R,T)$ gravity, taking into account a bulk viscous fluid, the Universe can experience an accelerated expansion behavior at the late times without CC problem.
\par
In Sect.~\ref{sec3}, taking the functionality of $f(R,T)$ as given in Sect.~\ref{sec2}, we show that a dissipative cosmic fluid with zero isotropic pressure can make the Universe to bear an admissible late time accelerated expansion. In this scenario the Universe evolves from an initial state with $w_{{\rm eff}}=(-1 + 2 n)/3$ and finally reaches an accelerated expansion state with $w_{{\rm eff}}=-1+[2n (n+1)]/[3 + n (-1 + 2 n)]$ in the late time regime. This is a novel result, since, even a dust-like cosmic matter provides accelerated expansion of the Universe achieving thus $w_{{\rm eff}}\approx-1$ and $q\approx-1$ in the late times. Also, we numerically obtain relevant quantities that already appear in the transport equation for the bulk viscosity under Eckart, Truncated Israel-Stewart (TIS) as well as Full Israel-Stewart (FIS) theories, consistent to our cosmological solutions. An important feature of these solutions is that for a cosmic fluid with $p=w\rho$, the bulk viscous stress is proportional to the energy density of fluid, i.e., $\Pi=w_{\Pi} \rho$ in which the coefficient of proportionality $w_{\Pi}$ is a simple function of both $w$ and $n$ parameters. Sect.~\ref{sec7} is devoted to the entropy production in this theory where we find that the rate of entropy production decreases while the Universe continuously expands. Finally, in Sect.~\ref{sec8} we summarize our discussion.
\section{Late time cosmological solutions: non-viscous fluid}\label{sec2}
In this section we investigate the late time behavior of the Universe which is filled by a non-viscous perfect fluid in the context of $f(R,T)$ theory of gravity. We suppose that contribution to the spatial curvature is negligible and obtain cosmological solutions for a perfect fluid with a barotropic EoS parameter, i.e., $p=w\rho$. The late time behavior of the Universe in the presence of a barotropic perfect fluid has so far been less studied in $f(R,T)$ gravity, hence, it would be advantageous to study it before proceeding to account for the viscosity effects. In this context the only cases which have thoroughly been considered are those models with vanishing pressure~\cite{shabani2013,shabani2014,shabani20172}. 

The $f(R,T)$ modified gravity is governed by the following action~\cite{harko2011}
\begin{align}\label{eq1}
S=\int \sqrt{-g} d^{4} x \left[\frac{1}{2\kappa^{2}} f(R,T)+L \right],
\end{align}
where $\kappa^{2}=8 \pi G$, $R$ indicates the Ricci scalar, $T$ denotes the trace of EMT and $L$ is the Lagrangian for matter sector of the action. Also, $f(R,T)$ is an arbitrary function of the Ricci scalar and the trace of EMT. Assuming that the matter Lagrangian depends only on the metric, the EMT is defined as
\begin{align}\label{eq2}
T_{\mu \nu}=g_{\mu \nu}L-2\frac{\partial L}{\partial g^{\mu \nu}}.
\end{align}
The corresponding field equation can then be obtained by metric variation of action~(\ref{eq1}), which gives
\begin{align}\label{eq3}
F(R,T) R_{\mu \nu}-\frac{1}{2} f(R,T) g_{\mu \nu}+\Big{(}g_{\mu \nu}
\square -\triangledown_{\mu} \triangledown_{\nu}\Big{)} F(R,T)=\Big{[}\kappa^{2}- {\mathcal F}(R,T)\Big{]} T^{\textrm{(m)}}_{\mu \nu}-{\mathcal F}(R,T)\Theta_{\mu \nu},
\end{align}
where we have used
\begin{align}
{\mathcal F}(R,T) \equiv \frac{\partial f(R,T)}{\partial T}~~~~\mbox{and}~~~~F(R,T) \equiv \frac{\partial f(R,T)}{\partial R}.\nonumber
\end{align}
Also the rank two tensor $\Theta_{\mu \nu}$ in the field equation~(\ref{eq3}) is defined  as
\begin{align}\label{eq4}
\Theta_{\mu \nu}\equiv g^{\alpha \beta} \frac{\delta T_{\alpha \beta}}{\delta g^{\mu \nu}}=-2T_{\mu \nu}+g_{\mu \nu}p,
\end{align}
where the matter Lagrangian $L=p$ has been supposed. We assume the geometry of the Universe is described by the flat FLRW metric
\begin{align}\label{eq5}
ds^{2}=-dt^{2}+a^{2}(t) \left [dr^{2}+r^{2}d\Omega^2\right ],
\end{align}
where $a(t)$ and $d\Omega^2$ indicate the scale factor and the line element on a unit two-sphere, respectively. From Eq.~(\ref{eq2}) the EMT is found as
\begin{align}\label{eq6}
T_{\mu\nu}=(\rho+p)u_\mu u_\nu + p g_{\mu\nu},
\end{align}
where $p$, $\rho$ and $u^\beta$ are the pressure, the density and the velocity four-vector of the cosmic fluid, respectively. Using  metric~(\ref{eq5}), the field equation~(\ref{eq3}) for theories with $f(R,T)=R+h(T)$ gives the modified Friedman equations, as follows
\begin{align}
&3H^{2}+\frac{1}{2}h(T)=\kappa^{2}\rho+h'(T)(\rho+p),\label{eq7}\\
&2\dot{H}=-\left[\kappa^{2}+h'(T)\right](\rho+p),\label{eq8}
\end{align}
where $H$ is the Hubble parameter, a dot shows derivative with respect to cosmic time and a prime denotes derivative with respect to the argument. Here, in order to consider minor deviations from GR, we have chosen the $f(R,T)$ function as a sum of Ricci scalar and a function of the trace of EMT. Combining Eqs.~(\ref{eq7}) and~(\ref{eq8}) leaves us with a simple equation which is served as the main equation and pictures the dynamics of the Universe. One then obtains
\begin{align}\label{eq9}
2\dot{H}+3H^{2}+\frac{1}{2}h(T)=-p,
\end{align}
where we have set the units so that $\kappa^{2}=1$. On the other hand, conservation of EMT tensor, i.e., $\nabla_{\mu}T^{\mu\nu}=0$, along with applying the Bianchi identity on the field equation~(\ref{eq3}) gives a very simple constraint on $h(T)$ function. This relation which plays an important role in our study can be derived as
\begin{align}\label{eq10}
\frac{1}{2}h'(T)=\rho h''(T).
\end{align}
Using the relation $T=-\rho+3p$ and the condition $h(0) = 0$, constraint~(\ref{eq10}) gives $p=(2n-1)\rho/3$ for the power-law form $h(T)=\alpha T^{n}$\footnote{To avoid complex values, we use $T\to -T$ for $w<1/3$ wherever $T$ being powered.}. For a barotropic perfect fluid with $p=w \rho$ the mentioned result leads to $w=(2n-1)/3$ which has also been reported in~\cite{sun2016}. Thus, one gets a relation between the free parameter $n$ and the EoS parameter $w$. Hence, the conservation of EMT in $f(R,T)=R+\alpha T^{n}$ gravity model reduces the number of free parameters. These type of models obey the standard behavior for the matter density, i.e., $\rho=\rho_{0} a^{-3(w+1)}$ where $\rho_{0}$ serves as the present value\footnote{In this paper, all present values are denoted by the subscription ``0".} of energy density. We note that the values of  Hubble constant and $\rho_{0}$ satisfy Eq.~(\ref{eq7}). In particular, for a pressureless perfect fluid we obtain $n=1/2$ which is a famous model discussed in the literature~\cite{shabani2013,shabani2014,shabani20172}.
\par
Next, we proceed to obtain Eq.~(\ref{eq9}) in terms of redshift parameter with the help of relations $H=-(z+1)^{-1}dz/dt$ and $\dot{H}(t)=-(z+1)H(z)H'(z)$. A straightforward calculation then gives
\begin{align}\label{eq11}
-2 (z+1) HH'+3 H^{2} +w \rho_{0} (z+1)^{3 (w+1)}+\frac{\alpha}{2} \left[(1-3 w) \rho_{0}\right]^{\frac{3 w+1}{2}}  (z+1)^{\frac{3}{2} (w+1) (3 w+1)}=0,
\end{align}
where $1+z=a_{0}/a$, $a_{0}=1$, $n=(3w+1)/2$, $T=(3w-1)\rho$ and $\rho=\rho_{0} a^{-3(w+1)}$ have been utilized. In the same way, we can also obtain the effective EoS defined as $w_{{\rm eff}}=p_{{\rm eff}}/\rho_{{\rm eff}}=-1-2\dot{H}/3H^{2}$ for which, using Eqs.~(\ref{eq7}) and~(\ref{eq8}) in the limit of high and low redshifts, we reach at the following results
\vspace{3mm}
\begin{align}\label{eq12}
-\frac{1}{3}<w<\frac{1}{3}~\Rightarrow~\left\{
\begin{array}{l}
w_{{\rm eff}}^{z\to\infty}=w,\\\\
w_{{\rm eff}}^{z\to-1}=\frac{3 w-1}{3 w^2+w+2}=-1+\frac{(w+1)(3 w+1)}{3 w^2+w+2}.
\end{array}
\right.
\end{align}

\vspace{3mm}Numerical considerations show that for the other values of $w$ parameter, the Universe evolves either from a state with $w_{{\rm eff}}^{z\to\infty}\to 0$ for $w<-1/3$ or ends to a final state with $w_{{\rm eff}}^{z\to-1}\to 0$ for\footnote{Note that $z\to\infty$ and $z\to -1$ correspond to $a\to 0$ and $a\to\infty$, respectively.} $w>1/3$ which obviously are non-physical. Therefore, we study cosmological solutions considering only the range $-1/3<w<1/3$ which is physically meaningful. For this range of EoS parameter, the Universe evolves from a state with $w_{{\rm eff}}=w$ to a state with $w_{{\rm eff}}=(3 w-1)/(3 w^2+w+2)$. Note that, the effective EoS parameter varies within the range $-1<w_{{\rm eff}}<0$ for the interval $-1/3<w<1/3$ at late times; with a nearly ultra-relativistic perfect fluid as the only involved matter content, the Universe mimics a pressureless matter dominated era in the late times, however, when $w=-1/3$ the Universe can reach a DE era with $w_{{\rm eff}}=-1$. In the case of a dust matter ($w=0$) in the early times, the conserved version of $f(R,T)=R+\alpha T^{n}$ predicts a DE dominated era with $w_{{\rm eff}}=-1/2$ at late times. This result has been also reported in Ref.~\cite{shabani2013} for $f(R,T)=g(R)+\alpha T^{1/2}$ models including different $g(R)$ functions.
\par
\begin{figure}[ht!]
\begin{center}
\epsfig{figure=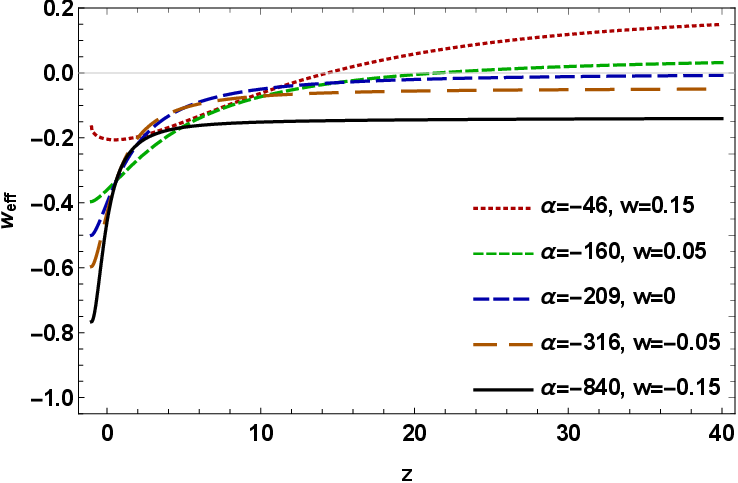,width=8.cm}\vspace{2mm}
\epsfig{figure=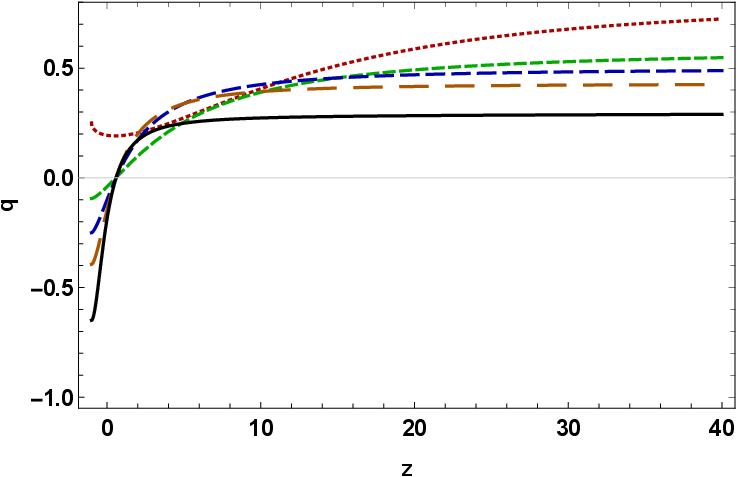,width=8.cm}
\caption{The effective EoS (left panel) and DP (right panel) predicted by the model $f(R,T)=R+\alpha T^n$ with a non-viscous barotropic perfect fluid and $n=(3w+1)/2$. Both diagrams have been plotted for the same values of model parameters with $H_{0}=67~kms^{-1}Mpc^{-1}$~\cite{aghanim2020}.}\label{f1}
\end{center}
\end{figure}
\begin{figure}[ht!]
\begin{center}
\epsfig{figure=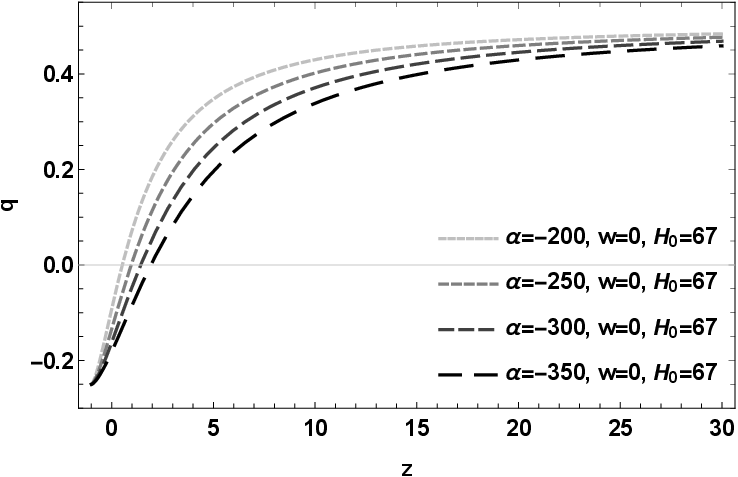,width=8.cm}\vspace{2mm}
\epsfig{figure=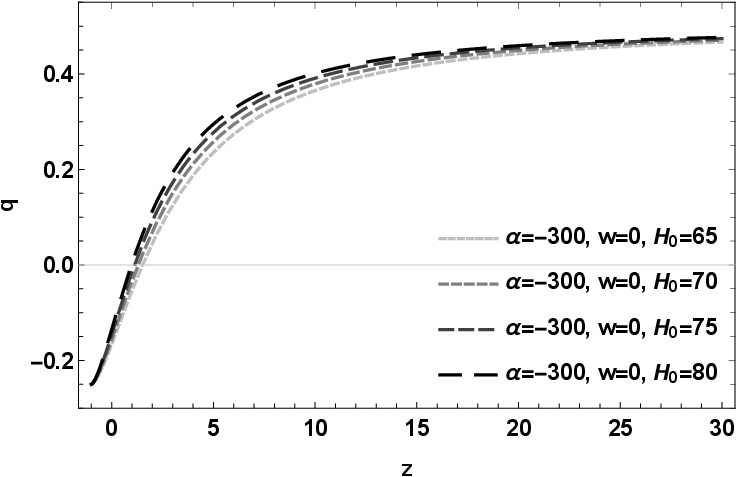,width=8.cm}
\caption{Evolution of DP with respect to redshift.}\label{f2}
\end{center}
\end{figure}

In Fig.~\ref{f1} the behavior of effective EoS as well as {DP} has been sketched\footnote{All figures have been drawn for the deceleration-acceleration transition redshift $z=0.6$~\cite{rania2015,muthukrishna2016,aghanim2020}.}. As is seen, the Universe evolves from states with $w_{{\rm eff}}=w$ to those with effective EoS obeying {the lower part of} Eq.~(\ref{eq12}). We also observe that consistent values of DP with astronomical data~\cite{camarena2020} can be obtained for negative values of $w$ within the range $-1/3<w<1/3$.  
Fig.~\ref{f2}, illustrates the effect of variations of parameter $\alpha$ and the Hubble constant $H_{0}$ on DP for $w$=0; different values of these parameters affect the present value of DP as well as the redshift at which the deceleration to acceleration transition takes place. Variations in $H_{0}$ makes minor effects on the evolution of the Universe.
\section{Late time cosmological solutions: viscous fluid}\label{sec3}
In this section, we consider effects of bulk viscosity and investigate how the Universe evolves under such assumption. To this aim, the bulk viscosity $\Pi$ is included in the total pressure as $\bar{p}=p+\Pi$, where $p$ is the isotropic pressure of the cosmic matter. The bulk viscous pressure is always a negative quantity in an accelerated expanding Universe~\cite{shogin2015,beesham2018,tamayo2022} and for the isotropic pressure we consider a linear relation between pressure and energy density, i.e., $p=w\rho$ with $w$ being the EoS parameter. The condition on conservation of EMT, Eq.~(\ref{eq10}), results in $\bar{p}=(2n-1)\rho/3$ in the case of $f(R,T)=R+\alpha T^{n}$ gravity, which in turn gives $\Pi=w_{\Pi}\rho$ with $w_{\Pi}=[(2n-1)/3]-w$. Therefore, the conservation of EMT makes the bulk viscosity to obey a barotropic EoS with the condition $n<(3w+1)/2$ which guarantees $\Pi<0$, always. Additionally, according to the Israel-Stewart theory, the bulk viscous stress $\Pi$ must be also smaller than the isotropic cosmic fluid pressure $p$ in magnitude. In other words, the condition $|\Pi|<p$ has to be also satisfied~\cite{chen2001}. This gives $|w_{\Pi}|<w$ or equivalently $1/2<n<(6w+1)/2$. Combining the mentioned conditions gives $1/2<n<(3w+1)/2$ provided $w>0$. However, as we will discuss in Sect.~\ref{sec7}, in an accelerated expanding Universe the total pressure $\bar{p}$, can also become negative provided that the processes of particle creation could occur in the Universe. Motivated by this fact, we disregard the constraint $|w_{\Pi}|<w$ and apply only the condition $\Pi<0$ which gives $n<(3w+1)/2$. For the specific case of $w=0$, this condition is satisfied when $n<1/2$. Moreover, when the bulk viscosity is turned off the condition $w=(2n-1)/3$ holds, the case which has been discussed in the previous section. Furthermore, in the current case by applying the conservation of EMT we {find} $\rho=\rho_{0}a^{-2(n+1)}$ for the matter density which means that, as a function of scale factor, its behavior is specified by the underlying gravitational model, instead of the EoS parameter of the cosmic fluid. In the particular case with $p=0=\bar\Pi$ or equivalently $n=1/2$, we recover $\rho=\rho_{0}a^{-3}$ formula which is consistent with the results previously mentioned.
\par
On the other hand, the bulk viscous stress $\Pi$ satisfies the transport equation~\cite{eckart1940,israel1979} given as
\begin{align}\label{eq13}
\Pi+\tau\dot{\Pi}=-3\zeta H-\frac{\epsilon}{2}\tau\Pi\left(3H+\frac{\dot{\tau}}{\tau}-\frac{\dot{\zeta}}{\zeta}-\frac{\dot{\mathbb{T}}}{\mathbb{T}}\right),
\end{align}
where $\zeta\geq0$ is the coefficient of bulk viscosity, $\tau\geq0$ is the relaxation time and $\mathbb{T}$ is temperature. By setting $\epsilon=1$ in Eq.~(\ref{eq13}) we obtain Full Israel-Stewart (FIS) theory and for $\epsilon=0$ Eq.~(\ref{eq13}) reduces to Truncated Israel-Stewart (TIS) model. The choice $\tau=0$ generates Eckart theory. In the presence of bulk viscosity Eqs.~(\ref{eq7})-(\ref{eq9}) still hold but for $\bar{p}$ instead of $p$. In this case, from Eq.~(\ref{eq9}) we obtain the following differential equation
\begin{align}\label{eq14}
-2 (z+1) HH'+3 H^{2} +\frac{\alpha}{2}  \left[2 ( n-1)\rho_{0}\right] ^{n} (z+1)^{2 n (n+1)} +\frac{(2 n-1)\rho_{0}}{3}  (z+1)^{2 (n+1)}=0,
\end{align}
where $\bar{p}=(2n-1)\rho/3$, $T=-\rho+3\bar{p}=2(n-1)\rho$ and $\rho=\rho_{0}a^{-2(n+1)}$. In the absence of bulk viscosity we have $w=(2n-1)/3$ and thus we recover Eq.~(\ref{eq11}) from Eq.~(\ref{eq14}). Our analyses show that cosmological solutions of the above equation are physically reasonable provided that $0<n<1$. In these cases the Universe evolves from the early states with $w^{{\rm early}}_{{\rm eff}}=(-1 + 2 n)/3$ to the late time eras with $w^{{\rm late}}_{{\rm eff}}=[3 (-1 + n)]/[3 + n (-1 + 2 n)]$.\\

Equation~(\ref{eq14}) governs the behavior of Hubble parameter with respect to redshift. Now, we first numerically simulate cosmological solutions of this equation and in the subsequent subsections we investigate consistency of the transport equation Eq.~(\ref{eq13}) with the obtained solutions. Note that the solutions of Eq.~(\ref{eq14}) are constrained via inequality $n<(3w+1)/2$. Also, specifying the parameters $w$ and $n$ gives $w_{\Pi}$ via the relation $w_{\Pi}=[(2n-1)/3]-w$ which in turn determines the behavior of bulk viscous stress $\Pi$. Meanwhile, the bulk viscosity $\Pi$ must also satisfy the transport equation which will be considered in the next subsections.\\

\begin{figure}[ht!]
\begin{center}
\epsfig{figure=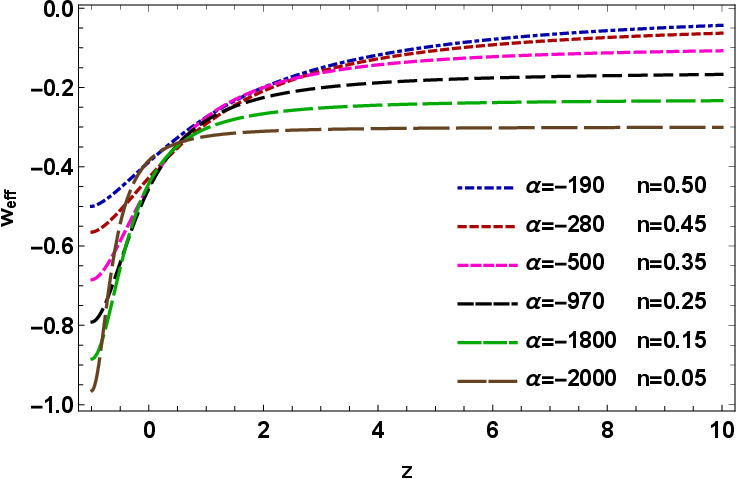,width=8.cm}\vspace{2mm}
\epsfig{figure=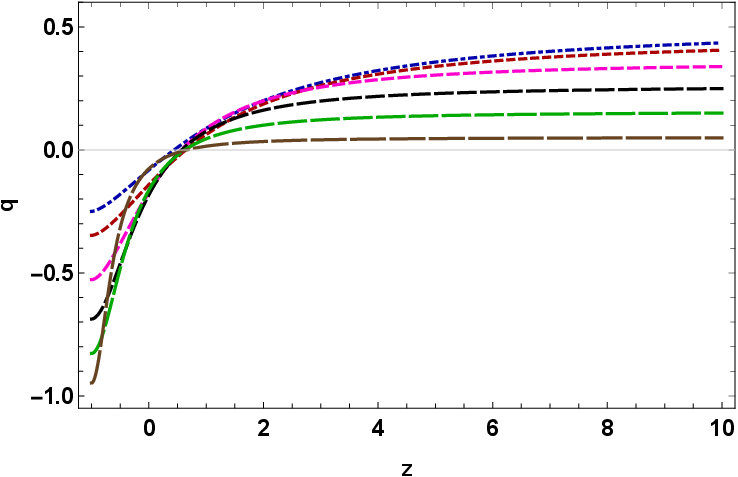,width=8.cm}\vspace{2mm}
\caption{Behavior of cosmological quantities for the model $f(R,T)=R+\alpha T^n$ including bulk viscosity effects. All panels are again provided for the same values of model parameters and $H_{0}=67~kms^{-1}Mpc^{-1}$ has been utilized. Here, $w=0.01$ corresponding to $n<0.515$ has been chosen.}\label{f3}
\end{center}
\end{figure}

The behavior of effective EoS and DP have been sketched in Fig.~\ref{f3} for $w=0.01$. These plots are generated for different values of the model parameters $\alpha$ and $n$. Here, in contrast to the cases for which the bulk viscosity is absent, the EoS parameter affects evolution of the Universe only through the relation $n<(3w+1)/2$. Interestingly, the bulk viscosity leads to different cosmological scenarios when a single cosmic fluid exists. In this regard, fixing the value of $w$, different scenarios for accelerated expansion at late times are predicted by the model, depending on the value of parameter $n$. 

In the previous section we saw that a pressureless perfect fluid (without viscosity effects) makes the Universe to reach a late time era with $q=-1/4$ (see the blue curve in Figs.~\ref{f1}), nevertheless, the presence of bulk viscosity allows for different values of DP at the late times depending on the model parameter $n$.
\par
\begin{figure}[ht!]
\begin{center}
\epsfig{figure=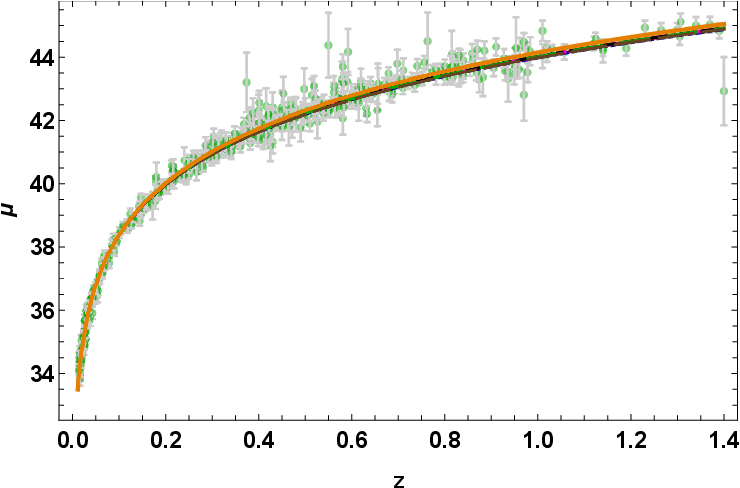,width=8.cm}
\caption{Behavior of the distance modulus $\mu(z)$ for the same model parameters as of Fig.~\ref{f3}. The yellow curve denotes the $\Lambda CDM$ distance modulus.}\label{mu}
\end{center}
\end{figure}

It is worthy of mention that for $n<0$ the Universe evolves from an initial state with $w^{{\rm early}}_{{\rm eff}}=0$ irrespective of the value of $n$ and reaches a late time state with $w^{{\rm late}}_{{\rm eff}}=[3 (-1 + n)]/[3 + n (-1 + 2 n)]$. While, all models with $n>1$, describe the Universe in a such a way that it evolves to a zero effective EoS parameter. The behavior of modulus $\mu(z)$ has been depicted in Fig.~\ref{mu} for $H_{0}=67 km s^{-1}Mpc^{-1}$ and different values of $\alpha$ and $n$ parameters. In this figure the yellow curve indicates the $\Lambda CDM$ distance modulus. We used the astronomical data from~\cite{amanullah2010,farooq2017} in Fig.~\ref{mu}.
 
In the following subsections, we first consider consistency of the above results with the most simple case, that is, Eckart theory for which $\tau=0$ holds, and then check out TIS theory with $\epsilon=0$, and the FIS equation including $\epsilon=1$ is finally discussed.
\subsection{Eckart theory ($\tau=0$)}\label{sec4}
\begin{figure}
\begin{center}
\epsfig{figure=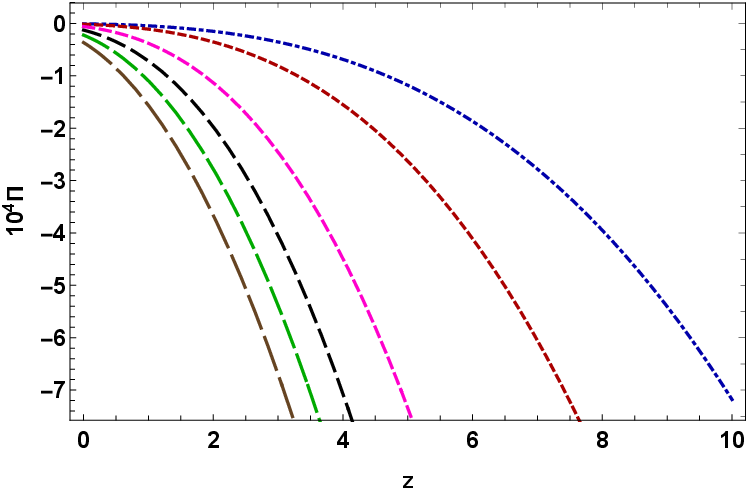,width=8.cm}\vspace{2mm}
\epsfig{figure=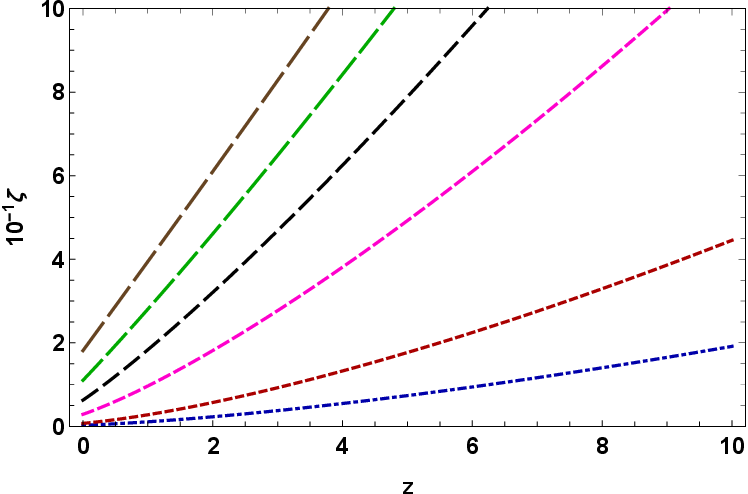,width=8.cm}
\caption{Evolution of the bulk viscosity $\Pi$ (left panel) and the $\zeta$ function (right panel) considering the Eckart equation for the same model parameters as of Fig.~\ref{f3}.}\label{f4}
\end{center}
\end{figure}

Eckart theory is described by setting $\tau=0$ in the transport equation~(\ref{eq13}), which gives
\begin{align}\label{eq15}
\Pi=-3\zeta H.
\end{align}
Since the conservation of EMT gives rise to a barotropic formulation for bulk viscosity, as discussed at the beginning of the current section, Eq.~(\ref{eq15}) gives the $\zeta$ function in terms of matter density and inverse of Hubble parameter. In fact, we obtain $\zeta=-\Pi/(3H)=-(w_{\Pi}/3)\rho H^{-1}$, where the Hubble parameter $H$ is the solution of Eq.~(\ref{eq14}) and $\rho=\rho_{0}(z+1)^{2(n+1)}$. In Fig.~\ref{f4} the behavior of $\zeta$ function and the bulk viscosity $\Pi$ is shown for $w=0.01$ and also the same values of model parameters as those of Fig.~\ref{f3}. Similar behavior for $\Pi$ has also been reported in~\cite{chen2001}.
Therefore, a compatible framework can be obtained in $f(R,T)=R+\alpha T^{n}$ gravity, as is demonstrated in Fig.~\ref{f3} and Fig~\ref{f4} as long as the Eckart equation (\ref{eq15}) does matter.
%
\subsection{Truncated Israel-Stewart (TIS) theory ($\epsilon=0$)}\label{sec5}
TIS theory is described by setting $\epsilon=0$ in Eq.~(\ref{eq13}) which reads
\begin{align}\label{eq16}
\Pi+\tau\dot{\Pi}=-3\zeta H,
\end{align}
where as mentioned at the beginning of Sect.~\ref{sec3} the bulk viscosity is $\Pi=w_{\Pi}\rho$ with $w_{\Pi}=[(2n-1)/3]-w$. Substituting the expression of bulk viscosity in Eq.~(\ref{eq16}) we find
\begin{align}\label{eq17}
\tau=\frac{1}{2(n+1)}\left[\frac{3}{w_{\Pi}\rho_{0}}(z+1)^{-2(n+1)}\zeta+H^{-1}\right].
\end{align}
The above equation shows a simple relation between $\tau$ and $\zeta$. If we suppose that the $\zeta$ function depends on the matter density in the form $\zeta=\beta\rho^{s}$, where $\beta>0$ and $s>0$ (see e.g.,~\cite{meng2007, brevik2008,jou2010}), Eq.~(\ref{eq17}) then gives
\begin{align}\label{eq18}
\tau=\frac{1}{2(n+1)}\left\{\frac{3\beta}{w_{\Pi}}\left[\rho_{0}(z+1)^{2(n+1)}\right]^{s-1}+H^{-1}\right\}.
\end{align}

\begin{figure}[ht!]
\begin{center}
\epsfig{figure=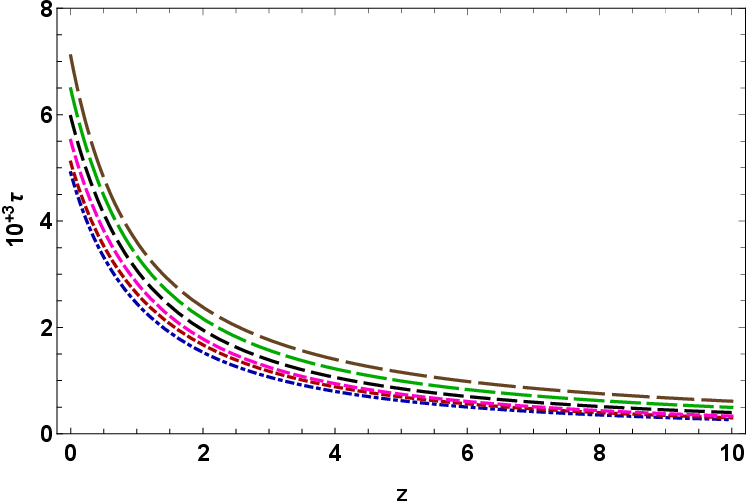,width=8.cm}\hspace{2mm}
\epsfig{figure=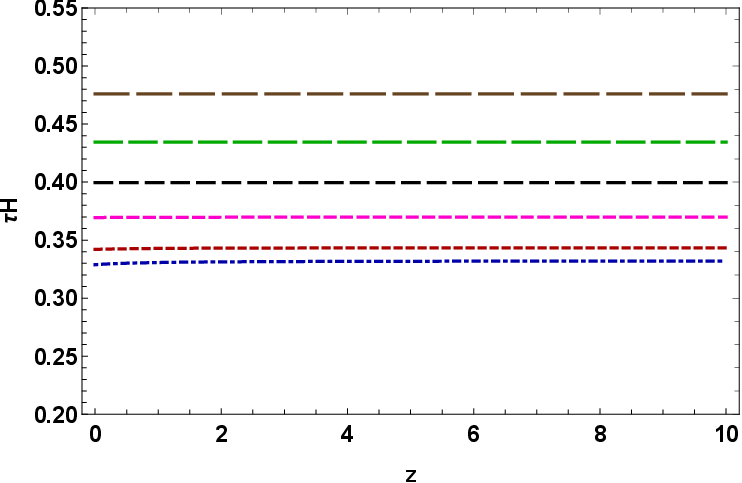,width=8.cm}
\caption{Evolution of relaxation time $\tau$ and $\tau H$ for $s=0.5$ and $\beta=1\times 10^{-4}$. These plots have been generated for the same values of model parameters as of Fig.~\ref{f3}. }\label{f5}
\end{center}
\end{figure}

The relaxation time $\tau$ grows from negligible values in the early times to maximum values in the late times as shown in Fig.~\ref{f5}. It is seen that, smaller values of parameter $n$ lead to slightly more growth for $\tau$. Therefore, a proper consistency between Eq.~(\ref{eq14}) and Eq.~(\ref{eq16}) can be obtained if we choose $\zeta=\beta\rho^{s}$. We have also plotted the behavior of $H\tau$ to check the condition $\tau H<1$. This condition guarantees that the timescale of physical interactions is less than the Hubble time. We see that the TIS transport theory satisfies this condition in the context of $f(R,T)$ gravity.
%
\subsection{Full Israel-Stewart theory ($\epsilon=1$)}\label{sec6}
\begin{figure}[ht!]
\begin{center}
\epsfig{figure=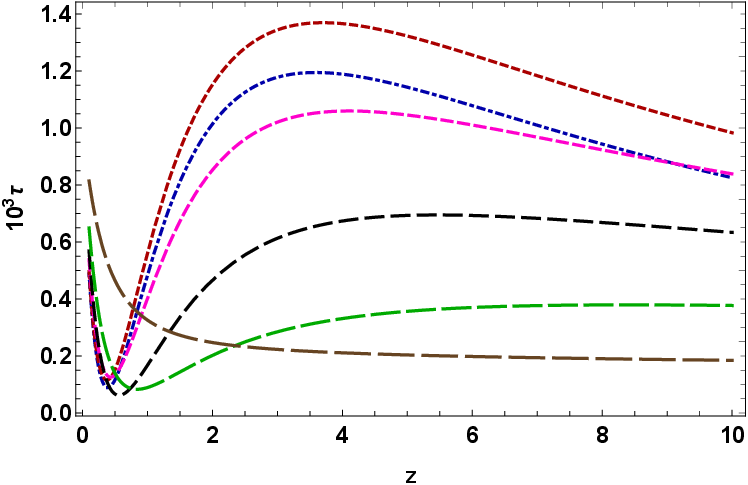,width=8.cm}\hspace{2mm}
\epsfig{figure=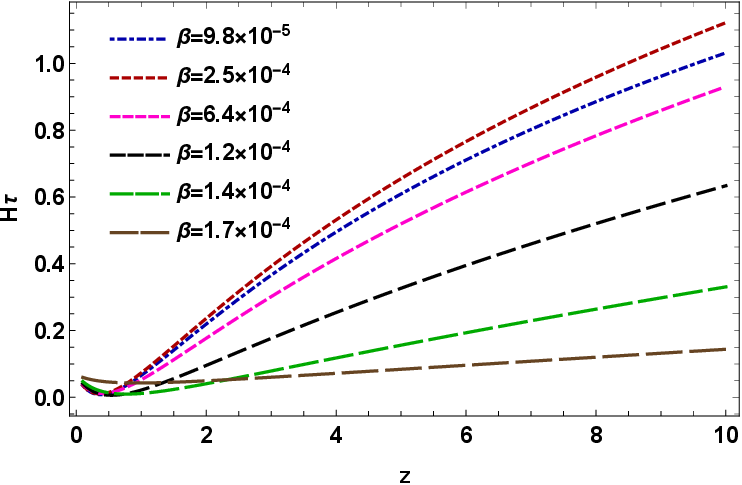,width=8.cm}
\caption{Evolution of relaxation time $\tau$ and $H\tau$ which are governed by the FIS transport equation. The same values of $\alpha$, $n$ and $H_0$ parameters as those of Fig.~\ref{f3} have been used. Also, we have set $s=0.5$, $\tau_{0}=1\times 10^{-3}$ and different values of $\beta$ parameter.}\label{f6}
\end{center}
\end{figure}

The most complete form of transport equation~(\ref{eq13}) is given by
\begin{align}\label{eq19}
\Pi+\tau\dot{\Pi}=-3\zeta H-\frac{1}{2}\tau\Pi\left(3H+\frac{\dot{\tau}}{\tau}-\frac{\dot{\zeta}}{\zeta}-\frac{\dot{\mathbb{T}}}{\mathbb{T}}\right).
\end{align}
Equation~(\ref{eq19}) includes three unknown quantities; relaxation time $\tau$, the coefficient of bulk viscosity $\zeta$ and temperature $\mathbb{T}$. Since, the number of equations is less than the number of unknowns, Eq.~(\ref{eq19}) can only be solved provided that two of the unknowns are determined. In order to deal the above equation we assume  $\zeta=\beta\rho^{s}$ for the relaxation time and $\mathbb{T}\propto \rho^{\frac{w}{2(w+1)}}$ (as will be obtained in Sec.~\ref{sec7}) for the temperature.
Using these assumptions in addition to the fact that bulk viscosity in the conserved version of $f(R,T)=R+\alpha T^{n}$ takes the form $\Pi= w_{\Pi}\rho$, we arrive at the following differential equation for the relaxation time
\begin{align}\label{eq20}
(z+1)\tau'-\left[3+2(n+1)\left(s+\frac{w}{2(w+1)}-2\right)\right]\tau-&\frac{18\beta}{2n-1-3w}(z+1)^{2(s-1)(n+1)}-2H^{-1}=0.
\end{align}
Equations~(\ref{eq14}) and~(\ref{eq20}) make a system of differential equations for the two unknown variables $\tau(z)$ and $H(z)$. In Fig.~\ref{f6} we have sketched numerical solution to the system~(\ref{eq14}) and~(\ref{eq20}) for different values of model parameters.
\\\\We conclude the present section with this result that under TIS theory in the presence of bulk viscosity, $f(R,T)=R+\alpha T^{n}$ gravity is capable of predicting an accelerated expansion picture for the evolution of the Universe. Under TIS theory, physical interactions occur within the timescales that is less than the Hubble time. However, for FIS theory their effects become more significant when the values of $n$ approach to zero.
\section{Temperature and entropy production}\label{sec7}
In the present section we discuss some thermodynamics properties of $f(R,T)=R+\alpha T^{n}$ gravity in the flat FLRW spacetime assuming the conservation of EMT. As discussed in~\cite{zimdah2001}, the process of particle creation allows for a large negative bulk viscous pressure so that the total pressure becomes negative, i.e., $\bar{p}<0$. Such a situation may occur in an accelerated expanding Universe~\cite{shogin2015}. This means that the constraint $n<(1+3w)/2$ (corresponding to $\Pi<0$) is the only condition which must be fulfilled. Hence, we let the total number of particles vary. We assume the number density is $\mathfrak{n}=N/V$ with $V=V_{0}a^{3}$, where $N$ is total particle number and $V$ denotes spatial volume. From the previous section we have $p=w\rho$, $\Pi=w_{\Pi}\rho$ with $w_{\Pi}=[(2n-1)/3]-w$ and $\bar{p}=(2n-1)\rho/3$ for the isotropic pressure and the bulk viscous stress, respectively, and $\bar{p}=p+\Pi$. For the 4-velocity vector field $u^{\beta}$, the flow of number density is defined as $N^{\mu}=\mathfrak{n}u^{\mu}$. The governing equations for the number density as well as conservation of EMT then read
\begin{align}
&\nabla_{\alpha}N^{\alpha}=\dot{\mathfrak{n}}+3H\mathfrak{n}=\Gamma\mathfrak{n},\label{eq21}\\
&\dot{\rho}+3(\rho+\bar{p})H=\dot{\rho}+2(n+1)\rho H=0.\label{eq22}
\end{align}

For $\Gamma>0$ ($\Gamma<0$) one has particle creation (annihilation). Using the mentioned assumptions, for the Gibbs law~\cite{maartens1996} one gets
\begin{align}\label{eq24}
d\sigma=d\left(\frac{\rho}{\mathfrak{n}}\right)+pd\left(\frac{1}{\mathfrak{n}}\right)=\frac{1}{\mathbb{T}\mathfrak{n}}\left[-\left(\frac{\rho+p}{\mathfrak{n}}\right)d\mathfrak{n}+d\rho\right],
\end{align}
where $\sigma$ denotes the entropy density. Using Eqs.~(\ref{eq21}) and~(\ref{eq22}) along with Eq.~(\ref{eq24}) we get an expression for the entropy production as~\cite{cruz2022}

\begin{align}\label{eq32}
\frac{d\sigma}{dt}=-\frac{1}{\mathfrak{n}\mathbb{T}}\Big[(1+w)\rho\Gamma+3\Pi H\Big].
\end{align}

Usually in dissipative processes the adiabatic (isentropic) particle production condition, $d\sigma/dt=0$, is adopted~\cite{zimdah2001}, which means the entropy per particle does not change. Under this situation the following result can be achieved, as

\begin{align}\label{eq33}
\Gamma=-\frac{3}{1+w}\frac{\Pi H}{\rho},
\end{align}

from which using $\Pi=w_{\Pi}\rho$ one gets

\begin{align}\label{eq34}
\Gamma=-\frac{3w_{\Pi}}{1+w}H.
\end{align}

Thus, in the conserved version of $f(R,T)=R+\alpha T^{n}$ theories the rate of particle production is proportional to the Hubble parameter under the adiabatic condition. Substituting the above result into Eq.~(\ref{eq21}) leads to a simple relation for $\mathfrak{n}$ as follows

\begin{align}\label{eq34a}
\mathfrak{n}=\mathfrak{n}_{0}(z+1)^{\frac{2 (n+1)}{w+1}}.
\end{align}

Now, we proceed  with calculating an expression for the temperature $\mathbb{T}$; The entropy density $\sigma$ satisfies the following integrability condition on $\mathfrak{n}$ and $\rho$
\begin{align}\label{eq25}
\left[\frac{\partial}{\partial\rho}\left(\frac{\partial \sigma}{\partial \mathfrak{n}}\right)_{\rho}\right]_{\mathfrak{n}}=\left[\frac{\partial}{\partial \mathfrak{n}}\left(\frac{\partial \sigma}{\partial\rho}\right)_{\mathfrak{n}}\right]_{\rho}.
\end{align}
If one supposes that temperature is a function of the number and energy densities, i.e., $\mathfrak{n}$ and $\rho$, the integrability condition leads to
\begin{align}\label{eq26}
\mathfrak{n} \frac{\partial\mathbb{T} }{\partial\mathfrak{n}}+(\rho+p) \frac{\partial\mathbb{T} }{\partial\rho}=\mathbb{T} \frac{\partial p}{\partial\rho},
\end{align}
whereby the above equation for the present model becomes
\begin{align}\label{eq27}
\mathfrak{n} \frac{\partial\mathbb{T} }{\partial\mathfrak{n}}+(1+w)\rho \frac{\partial\mathbb{T} }{\partial\rho}=w\mathbb{T}.
\end{align}
The chain rule for the first term in the above equation reads
\begin{align}\label{eq27a}
\mathfrak{n} \frac{\partial\mathbb{T} }{\partial\mathfrak{n}}=\mathfrak{n} \frac{\partial\mathbb{T} }{\partial\rho}\frac{\partial\rho}{\partial a}\frac{\partial a}{\partial t} \left[\frac{\partial \mathfrak{n}}{\partial t}\right]^{-1},
\end{align}

which leads to

\begin{align}\label{eq27b}
\mathfrak{n} \frac{\partial\mathbb{T} }{\partial\mathfrak{n}}=(1+w) \rho\frac{\partial\mathbb{T} }{\partial\rho},
\end{align}

where $\rho=\rho_{0}a^{-2(n+1)}$ and $\mathfrak{n}=\mathfrak{n}_{0}a^{-\frac{2 (n+1)}{w+1}}$ have been used. Therefore, combining Eqs.~(\ref{eq27}) and~(\ref{eq27b}) we obtain

\begin{align}\label{eq27c}
 \frac{\partial\mathbb{T} }{\mathbb{T}}=\frac{w}{2(w+1)}\frac{\partial\rho }{\rho},
\end{align}

which gives the following expression for the temperature

\begin{align}\label{eq27d}
\mathbb{T}=\mathbb{T}_{0}\rho^{\frac{w}{2(1+w)}.}=\mathbb{T}_{0}a^{-\frac{w(n+1)}{w+1}}.
\end{align}

We come to the conclusion that in $f(R,T)=R+\alpha T^{n}$ gravity in the absence of viscosity the relevant thermodynamics quantities depend on the scale factor as follows

\begin{align}\label{eq35}
\rho\propto a^{-3},~~~~~~~\mathfrak{n}\propto a^{-3},~~~~~~\mathbb{T}\propto constant,
\end{align}
where the conservation of EMT as well as the adiabatic condition are assumed.

%
\section{Concluding Remarks}\label{sec8}
In this work we have investigated late time cosmological solutions of $f(R,T)=R+\alpha T^{n}$ gravity in a flat FLRW spacetime in the presence of both a barotropic perfect fluid as well as an isotropic cosmic fluid with a bulk viscosity. The issue has been considered by the condition that the EMT is conserved. This model is chosen since it introduces minor deviations from GR. We found that when a perfect fluid with EoS $p=w\rho$ serves as the cosmic matter, the Universe evolves from an initial state with effective EoS parameter $w_{{\rm eff}}=w$ to a final stage for which $w_{{\rm eff}}=-1+L(w)$ (or in terms of DP, $q=-1+3L(w)/2$) where $L(w)=(w+1)(3 w+1)/(3 w^2+w+2)$.  In these cases, we have $-1<w_{{\rm eff}}<0$ in the late times. Analytical and numerical inspections of equations of motion show that such a scenario is valid only for $-1/3<w<1/3$. Otherwise, the Universe either begins from a situation with $w_{{\rm eff}}=0$ for $w<-1/3$ or evolves toward a state with $w_{{\rm eff}}=0$ for $w>1/3$. In the particular case of a pressureless perfect fluid, the Universe reaches a late time DE era with $w_{{\rm eff}}=-1/2$ or equivalently $q=-1/4$. This output has been reported in~\cite{shabani2013} via a dynamical system approach. However, a cosmic matter mimicking a quintessence can give rise to more observationally justified results for DP. Also, the effects of different values of the model parameter $\alpha$ and Hubble constant $H_{0}$ on the evolution of DP has been examined. We observed that variation of $H_{0}$ has a minor effect comparing with the coupling constant $\alpha$. In both cases, only the redshift of deceleration to acceleration phase transition and also the present value of DP changes. 

We have next considered evolution of the Universe which is occupied by a cosmic fluid endowed with the isotropic pressure $p=w\rho$ and a bulk viscosity $\Pi$ so that the total pressure of matter content is $\bar{p}=p+\Pi$. With this new assumption, by applying the conservation of EMT which we obtained barotropic formulation for the total and the bulk viscous pressures, as $\bar{p}=(2n-1)\rho/3$ and $\Pi=([(2n-1)/3]-w) \rho$. We observed that the total pressure depends only on the model parameter $n$ contrary to the condition in which $\Pi$ is zero. Assuming that the total number of particles in the Universe is not preserved the result we obtained is that, physically viable models include $0<n<1$ with the restriction $n<(3w+1)/2$ in order to ensure $\Pi<0$ and $n<(6w+1)/2$ to satisfy the condition $|\Pi|<p$. Therefore, for each value of $w$ we arrive at a range for the values of parameter $n$, i.e., $n< n_{max}$ where $n_{max}$ is determined via the mentioned restriction. This helps us to choose the best value for parameter $n$ comparing to the observational data. We obtained numerical simulations for some cosmological quantities. The diagrams show that an acceptable description of the Universe can be found; even for negligible isotropic pressure of the cosmic fluid, the Universe evolves to a final state with reasonable values of DP. Proper relaxation time and the coefficient of bulk viscosity has been found by considering the transport equation under Eckart and Israel-Stewart theories, for the solution which obtained from the Friedman equations. We found that the coefficient of bulk viscosity tends to grow by increasing the redshift and conversely the relaxation time decreases as the redshift increases. Finally, considering the isotropic condition for particle production we derived power-law rules for the number density of particles as well as the temperature which appears in the complete form of the Israel-Stewart theory.
\par
We conclude our discussion with the final statement that $f(R,T)=R+\alpha T^{n}$ models may present a better description for the evolution of the Universe when the bulk viscosity is included. The mentioned model has been investigated in~\cite{shabani2013,shabani2014} for $n=1/2$ which corresponds to existence of only the pressureless matter. Comparing the results of~\cite{shabani2013,shabani2014} with the present outputs, we find that viscous interactions may lead to a late time accelerated expansion without having to introduce DE. In this report we tried to address such a possibility.

\section*{Data availability}
No data was used for the research described in the article. 


\section*{Acknowledgement}
This work has been supported by the Universidad de Santiago de Chile, USACH, through Proyecto DICYT N$^{\circ}$ 042131CM (N.C.), Vicerrector\'ia de Investigaci\'on, Desarrollo e Innovaci\'on.


\end{document}